\documentclass[showpacs,prl,amsmath,amssymb]{revtex4}
\usepackage{graphicx}
\begin{document}

\title{Surface Plasmon Instability Leading to Emission of Radiation}

\author{Godfrey Gumbs$^{1,2}$, Andrii Iurov$^{1}$, Danhong Huang$^{3}$, and Wei Pan$^{4}$}
\affiliation{$^{1}$Department of Physics and Astronomy, Hunter College of the
City University of New York, 695 Park Avenue, New York, NY 10065, USA\\
$^{2}$ Donostia International Physics Center (DIPC),
P de Manuel Lardizabal, 4, 20018 San Sebastian, Basque Country, Spain\\
$^{3}$Air Force Research Laboratory, Space Vehicles Directorate, Kirtland Air Force Base, NM 87117, USA\\
$^{4}$Sandia National Laboratory, Albuquerque, NM 87185, USA}

\begin{abstract}
We propose a new energy conversion approach from a dc electric field to a terahertz wave   based 
on  hybrid semiconductors by combining
two-dimensional (2D) crystalline layers and a thick conducting material with  possible applications
as a source of coherent radiation. The hybrid nano-structure may consist  of a single or pair
of sheets of graphene, silicene or a 2D electron gas as would occur at a semiconductor
hetero-interface. When an electric current is passed through a layer, we discover that the
low-frequency plasmons may become unstable beyond a critical wave vector $q_c$. However, there
is no instability for a single driven layer far from the conductor and the instability of an isolated pair of 2D layers
occurs only at ultra long wavelengths. To bring in frequency agility for this spontaneous radiation,
we manipulate the surface-plasmon induced instability, which leads to the emission of radiation (spiler),
to occur at shorter wavelengths by choosing the conductor electron density, layer separation, distances of layers from the
conductor surface and the driving-current strength. Applications of terahertz radiation from spiler
for chemical analysis, security scanning, medical imaging and telecommunications are expected.
\end{abstract}


\maketitle

Possible sources of terahertz (THz) radiation have been investigated for
several years now. These frequencies cover the electromagnetic (EM) spectrum
lying between microwave and far-infrared. By epitaxially growing layers of
different semiconductors (including GaAs, GaAs/AlGaAs, InAs/InGaAs),
multiple quantum-well layers, which emit high-power THz radiation across a wide frequency range,
have been fabricated. The work reported so far covers ultra-long wavelength emission,
phase/mode-locking, multiple color generation, photonic crystal structures, and
improved laser performance with respect to both maximum operating temperature
and peak output power.
It was predicted by Kempa, et al.\,\cite{Bakshi} (see also Ref.\,[\onlinecite{AI1}]) that when
a current is passed through a stationary electron gas, the
Doppler shift in response frequency of this two-component plasma leads to a spontaneous generation of
plasmon excitations at ultra-long wavelengths and subsequent Cherenkov radiation\,\cite{AI4star}
at sufficiently high carrier drift velocities. For their model, the process is irreversible
based on the lack of time reversal symmetry (Onsager's principle of microscopic reversibility)\,\cite{Onsager}.
Similar conclusions are expected for monolayer graphene which is
characterized by massless Dirac fermions where the energy dispersion is linear in the
wave vector ${\bf k}_\parallel$ or a nanosheet of silicene consisting of silicon atoms,
which has been synthesized\,\cite{silicene}.
In the same group of the periodic table with graphene, silicene is predicted to exhibit similar
electronic properties. Additionally, it has the advantage over graphene in
its compatibility with Si-based device technologies.
\medskip

The role played by plasma excitations in the THz response of low-dimensional microstructures has received considerable attention\,\cite{4,5,6,7,8,9,10,11,12,13}.
Plasmon modes of quantum-well transistor structures with frequencies in the THz range may be excited  with the use of far-infrared (FIR) radiation and other means\,\cite{Tso}. 
A split grating gate design  has been found to significantly enhance FIR response\,\cite{2,3,gg1,gg2}. 
Under this scheme, however, the stimulated EM radiation require either a population inversion\,\cite{laser} or a quantum coherence\,\cite{LWI}, or a condensation\,\cite{polar}.
The EM radiation can also be generated by transferring energy from optical field to another\,\cite{spaser}.
\medskip

Here, we explore a new energy conversion approach, i.e., from an applied dc electric field to an optical field based on a current-driven induced instability.
The primary objective of the Letter is to expand the materials platform by exploiting the functionality
of a composite nano-system consisting of a thick conductor (including a heavily-doped semiconductor)
which is Coulomb coupled to 2D layered materials. The Coulomb coupling of the plasmons in a layer to the surface plasmon on the conductor results in a surface-plasmon instability
that leads to the emission of radiation (spiler).
The predicted tunable spiler radiation relies on a current-induced plasmon instability and comes after the plasmon grows in the time domain at a rate which is determined by the surface-plasmon frequency,
the 2D layer separation, the distance of the 2D layers from the conducting surface and the driving-current strength.
\medskip

In our formalism, we consider a nano-scale system consisting
of a pair of 2D layers and a thick conducting  material. The
layer may be monolayer graphene or a 2DEG such as a
semiconductor inversion layer or high electron mobility transistor.
The graphene layer may have a gap, thereby extending the flexibility of the composite system that
also incorporates a thick conducting layer.
\medskip

In our notation, the structure has a double layer located
at $z=a_1$ and $z=a_2$  ($0<a_1<a_2$)  interacting with each other
as  well as the semi-infinite system with its surface lying
in the $xy$-plane at $z=0$. The longitudinal  excitation spectra
of allowable modes will be determined from a knowledge
of the frequency-dependent non-local dielectric function
$ \epsilon ({\bf r},{\bf r}^\prime;\omega)$ which depends
on the position coordinates ${\bf r}$, ${\bf r}^\prime$ and
frequency $\omega$. Alternatively, the normal modes correspond to the resonances
of the inverse dielectric function
$K({\bf r},{\bf r}^\prime;\omega)$, satisfying
$\int d{\bf r}^\prime\,K({\bf r},{\bf r}^\prime;\omega)\,\epsilon({\bf r}^\prime,{\bf r}^{\prime\prime};\omega)
=\delta({\bf r}-{\bf r}^{\prime \prime})$.   The significance of
$K({\bf r},{\bf r}^\prime;\omega)$ is that it embodies many-body effects\,\cite{AI2} through
screening by the medium of an external potential $U({\bf r}^\prime;\omega)$ to
produce an effective potential $V({\bf r};\omega)=\int d{\bf r}^\prime\,K({\bf r},{\bf r}^\prime;\omega)\,
U({\bf r}^\prime;\omega)$. The self-consistent
field equation for $K({\bf r},{\bf r}^\prime;\omega)$ is
in integral form, after Fourier transforming parallel to the $xy$-plane and
suppressing the in-plane wave number $q_\parallel$ and frequency $\omega$, leading to

\begin{equation}
K(z_1,z_2)= K_{SI}(z_1,z_2) - \sum_{j=1}^{2}\int_{-\infty}^\infty dz^\prime\int_{-\infty}^\infty
dz^{\prime\prime}\,K_{SI}(z_1,z^\prime)\,
\alpha_{2D;j}(z^\prime,z^{\prime\prime})\,K(z^{\prime\prime} ,z_2)\ .
\label{e1}
\end{equation}
Here, the polarization function for the 2D structure is given by

\begin{equation}
\alpha_{2D;j}(z^{\prime}, z^{\prime\prime})=\int_{-\infty}^\infty dz^{\prime\prime\prime}\,
v(z^\prime-z^{\prime\prime\prime})\,D_j(z^{\prime\prime\prime},z^{\prime\prime})\ ,
\label{e2}
\end{equation}
where $v(z-z^\prime)=(2\pi e^2/\epsilon_sq_\parallel)\,\exp(-q_\parallel|z-z^\prime|)$, $\epsilon_s=4\pi\epsilon_0\epsilon_r$, and the 2D response function obeys
$D_j(z^{\prime\prime\prime},z^{\prime\prime})=\Pi_{2D;j}^{(0)}(q_\parallel,\omega)\,
\delta(z^{\prime\prime\prime}-a_j)\,\delta(z^{\prime\prime}-a)$
with $\Pi_{2D;j}^{(0)}(q_\parallel,\omega)$ as the single-particle in-plane response.  Upon
substituting this form of the polarization function for the monolayer into
Eq.\,(\ref{e1}), we have

\begin{equation}
K(z_1,z_2)= K_{SI}(z_1,z_2)-\sum_{j=1}^{2}\,
\Pi_{2D;j}^{(0)}(q_\parallel,\omega)\int_{-\infty}^\infty  dz^\prime\,K_{SI}(z_1,z^\prime)\,v(z^\prime-a_j)\,K(a_j,z_2)\ .
\label{eq:GG}
\end{equation}
We now set $z_1=a_1$  and $z_1=a_2$  in turn in Eq.\,(\ref{eq:GG}) and
solve simultaneously the pair of equations for $K(a_1,z_2)$ and $K(a_2,z_2)$ to obtain

\begin{equation}
\left[
\begin{matrix} K(a_1,z_2)\cr
K(a_2,z_2)\cr
\end{matrix}
\right]=\frac{1}{S_c^{(2)}(q_\parallel,\omega)} \tensor{{\cal M}} (q_\parallel,\omega)
\left[
\begin{matrix} K_{SI}(a_1,z_2)\cr
K_{SI}(a_2,z_2)\cr
\end{matrix}
\right]\ ,
\label{eq:GGDH}
\end{equation}
where $S_c^{(2)}(q_\parallel,\omega)=\mbox{Det}[\tensor{{\cal M}} (q_\parallel,\omega)]$ with the coefficient matrix given by

\begin{eqnarray}
&& \tensor{{\cal M}}(q_\parallel,\omega)=
\nonumber\\
&& \left[\begin{array}{cc}
1+\Pi_{2D;2}^{(0)}(q_\parallel,\omega)\int\limits_{-\infty}^\infty dz^\prime\,K_{SI}(a_2,z^\prime)\,v(z^\prime-a_2) &
-\Pi_{2D;2}^{(0)}(q_\parallel,\omega)\int\limits_{-\infty}^\infty dz^\prime\,K_{SI}(a_1,z^\prime)\,v(z^\prime-a_2)  \cr
-\Pi_{2D;1}^{(0)} (q_\parallel,\omega)\int\limits_{-\infty}^\infty dz^\prime\,K_{SI}(a_2,z^\prime)\,v(z^\prime-a_1) &
1+\Pi_{2D;1}^{(0)}(q_\parallel,\omega)\int\limits_{-\infty}^\infty dz^\prime\,K_{SI}(a_1,z^\prime)\,v(z^\prime-a_1)
\end{array}
\right]\ .
\label{eq:M}
\end{eqnarray}
In our numerical calculations, we shall use
$K_{SI} (z,z^\prime)$ given in Eq.\,(30) of Ref.\,[\onlinecite{Horing}].
Substituting the results for $K_{SI}(a_1,z_2)$ and $K_{SI}(a_2,z_2)$ into Eq.\,(\ref{eq:GGDH}),
we obtain the complete inverse dielectric function for a pair
of 2D conducting planes interacting with each other and a semi-infinite conducting material.
The plasmon excitation frequencies are determined by the zeros of
$S_c^{(2)}(q_\parallel,\omega)$. Furthermore, the effect of the inverse dielectric
function for the semi-infinite structure  $K_{SI} (z,z^\prime;q_\parallel,\omega)$
leads to coupling between the two layers and of each layer with the
bulk and surface of the neighboring material. As a matter of fact, our result
for the plasmon dispersion relation generalizes that obtained by Das Sarma and
Madhukar\,\cite{DasSarma, Ai3Pohl, Polini} for a biplane. We obtain in the local limit\,\cite{Horing}

\begin{eqnarray}
S_c^{(2)}(q_\parallel,\omega) &=&\left\{1+\frac{2\pi e^2}{\epsilon_sq_\parallel}\,\Pi_{2D;2}^{(0)}(q_\parallel,\omega)
\left[1+e^{-2q_\parallel a_2}\,\frac{\omega_p^2}{2\omega^2-\omega_p^2}\right]\right\}
\nonumber\\
&\times&\left\{1+\frac{2\pi e^2}{\epsilon_sq_\parallel}\,\Pi_{2D;1}^{(0)}(q_\parallel,\omega)
\left[1+e^{-2q_\parallel a_1}\,\frac{\omega_p^2}{2\omega^2-\omega_p^2}\right]\right\}
\nonumber\\
&-&\left(\frac{2\pi e^2}{\epsilon_sq_\parallel}\right)^2
\Pi_{2D;1}^{(0)}(q_\parallel,\omega)\,\Pi_{2D;2}^{(0)}(q_\parallel,\omega)
\left[e^{-q_\parallel|a_1-a_2|}+e^{-q_\parallel(a_1+a_2)}\,\frac{\omega_p^2}{2\omega^2-\omega_p^2}\right]^2\ .
\label{eq:biplane}
\end{eqnarray}
\medskip

\begin{figure}[t]
\centering
\includegraphics[width=0.45\textwidth]{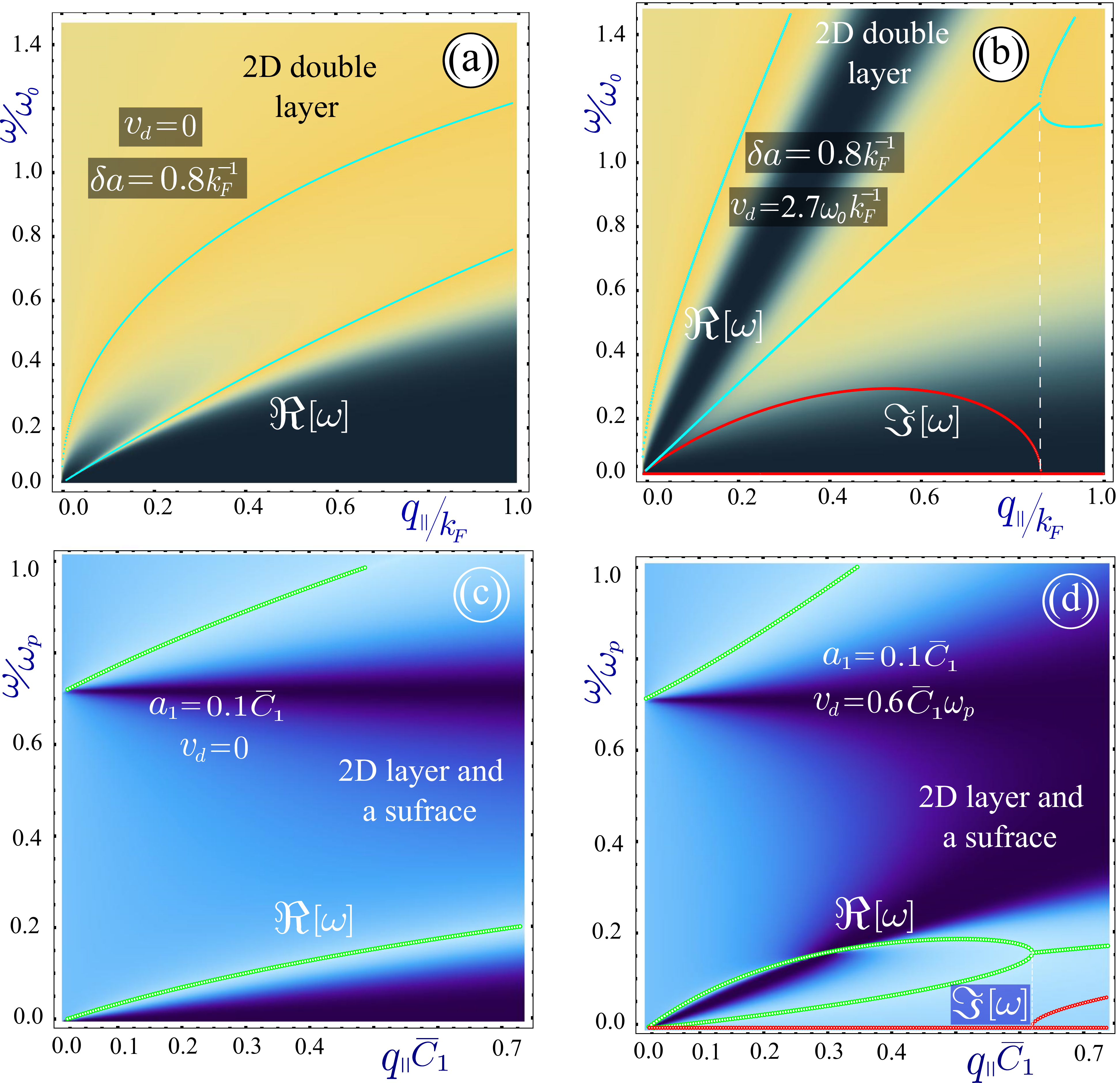}
\caption{(Color online) Complex frequencies yielding the plasmon dispersion
(real part $\mathfrak{R}[\omega]$) and inverse  growth rate (imaginary part $\mathfrak{I}[\omega]$)
for a pair of free-standing 2D layers with separation $\delta a$
[panels $(a)$ and $(b)$] and for the case of one layer Coulomb coupled to a
semi-infinite conducting substrate [panels $(c)$ and (d)].
The frequency unit in $(a)$ and $(b)$ is $\omega_0=\sqrt{2\pi e^2 C_1 k_F/\epsilon_s}$,
while the wave vector $q_\parallel$ is measured in units of the Fermi wave vector $k_F$.
In the case of two layers, the instability domain ranges from the origin to
a certain value of $q_\parallel$ at the bifurcation point, i.e., the ultra-long wavelength region. The position of this bifurcation point
depends on $\delta a$ and the drift velocity $v_d$.
The instability domain changes drastically in the presence of
a surface, as demonstrated in the two lower panels $(c)$ and $(d)$ with
$q_c$ shifted from zero to shorter wavelengths. Here, the current is passed through the 2D layer in panel $(d)$,
and it is passed through the bottom 2D layer in panel $(b)$. The carrier density,
temperatures or doping densities are chosen such that $\bar{C}_2=1.2\bar{C}_1$ for all cases.}
\label{FIG:1}
\end{figure}
Setting $a_1=a$ and letting $a_2\to\infty$ in Eq.\,(\ref{eq:M}), the off-diagonal
matrix elements tend to zero and the element in the first row and first column reduces
to unity. Subsequently, the dispersion equation for a single layer interacting with
the substrate is given by the zeros of the matrix element in the second row and
second column. Using the long-wavelength limit ($q\ll k_F$), we find
$\Pi_{2D;j}^{(0)} (q_\parallel,\omega) \approx -C_jq_\parallel^2/\omega^2$.
For a 2DEG, we have $C=n_{2D}/m_{2D}^\ast$;
for doped graphene, we have
$C=(2\mu/\pi\hbar^2)\left[1-(\Delta^2/\mu^2)\right]$,
where $\mu$ is the chemical potential and $\Delta$ is the gap
between valence and conduction bands; for intrinsic graphene whose plasmon excitations
are induced by temperature, $C=(2\ln 2)\,k_BT/\pi\hbar^2$\,\cite{SDSLi}.
Consequently, we find the plasmon frequency as follows\,\cite{NJMH}:
$\omega^2=K_1\pm\sqrt{K_2}$ with $K_1$ and $ K_2$ defined by
$K_1=\overline{C}q_\parallel\omega_p^2/2+\left(\omega_p/2\right)^2$ and
$K_2=\overline{C}q_{\parallel}\omega_p^4\exp(-2q_\parallel a)/2+(\omega_p/2)^4(1-2\overline{C}q_{\parallel})^2$, where $\overline{C}=2\pi e^2C/(\epsilon_s\omega_p^2)$.
Additionally, within this long-wavelength limit, these expressions yield the plasmon excitation frequencies
$\omega_1/\omega_p\backsimeq q_\parallel\sqrt{2\overline{C}a}$ and
$\omega_2/\omega_p\backsimeq 1/\sqrt{2}+\overline{C}q_{\parallel}/\sqrt{2}$
which are both linear in $q_\parallel$ and unlike the $\sqrt{q_\parallel}$-dependence for free-standing
graphene or the 2DEG\,\cite{wunch,pavlo,7+,8+,9+,10+}.
\medskip

\begin{figure}[t]
\centering
\includegraphics[width=0.45\textwidth]{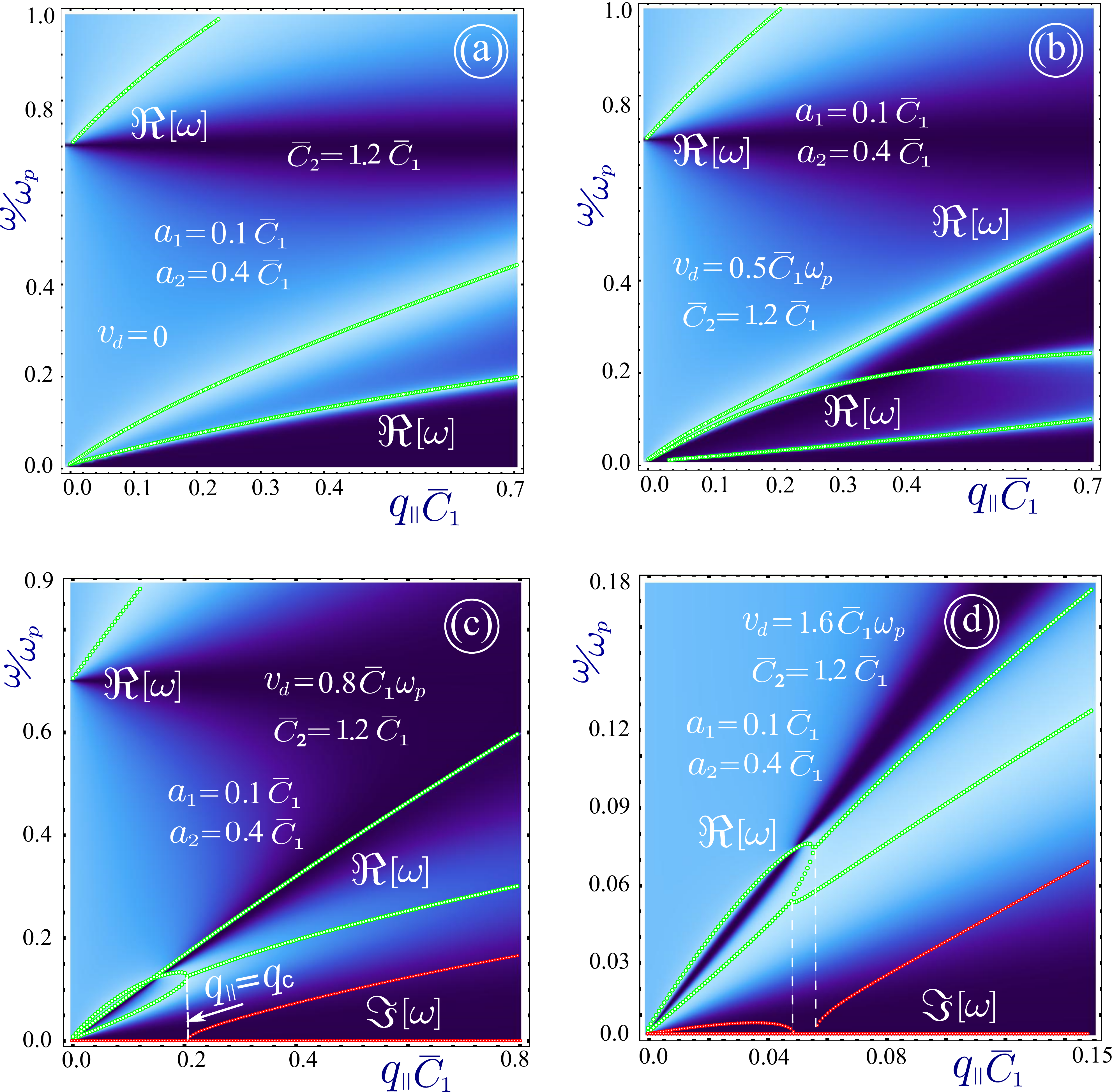}
\caption{(Color online) Plasmon dispersion relation and inverse growth
rate for a pair of 2D layers and a semi-infinite conducting medium.
The plasma frequency for the semi-infinite medium is $\omega_p$.
The layers are located at $a_1=0.1\bar{C}_1$ and $a_2=0.4\bar{C}_1$
with respect to the surface. Each panel corresponds to a different drift velocity
with $v_d/\omega_p=0,\,0.5\bar{C}_1,\,0.8\bar{C}_1$
and $1.6\bar{C}_1$. Here, the current is passed through the bottom 2D layer.
Panel $(a)$ with $v_d=0$ corresponds to the solutions in
Eq.\,(\ref{solutions}). Panel $(b)$ shows that for a small $v_d$ an additional
plasmon branch appears, but all solutions are stable.
The carrier concentrations, chemical potentials or temperatures in the layers are such that
$\bar{C}_2=1.2\bar{C}_1$ for all cases. Either of the two lowest
plasmon branches might become unstable, depending on $v_d$. The Rabi-type splitting
of the plasmon excitation branches by the external electric field is attributed to
quasiparticles with different excitation energies for the same wavelength. After the
Rabi ``loop" closes at $q_c$ in $(c)$, the lowest branch becomes unstable with finite imaginary
part illustrated by the red curve. In $(d)$, the lowest plasmon branch has an instability for two
separate ranges of wave vector.}
\label{FIG:2}
\end{figure}
In Ref.\,\cite{1+}, it was demonstrated that the plasmon excitations in graphene has a linear
dispersion rather than a square root dependence on the wave vector. This startling result
came as a surprise because theoretical calculations on free-standing graphene clearly
do not predict a  linear dependence in the long-wavelength limit.
As a matter of fact, this linear dependence of plasmon frequency  on wave vector was attributed
to local field corrections to the random-phase approximation.
In our notation, $\overline{C}_j=2\pi e^2C_j/(\epsilon_s\omega_p^2)$ for $j=1,2$.
The spectral function yields  real frequencies. A plane interacting with
the half-space has two resonant modes. Each pair of 2D layers interacting in isolation
far from the semi-space medium supports a symmetric and an
antisymmetric mode\,\cite{DasSarma}. In the absence of a driving current, the analytic solutions for the plasmon modes of a
pair of 2D layers that are Coulomb coupled to a half-space are given by

\begin{eqnarray}
\label{solutions}
&& \frac{\Omega_1(q_\parallel)}{\omega_p}=1/\sqrt{2}+q_\parallel(\overline{C}_1+\overline{C}_2)/\sqrt{2}+\mathcal{O}[q_\parallel^2]\ ,
\nonumber\\
&& \frac{\Omega_2(q_\parallel)}{\omega_p}=q_\parallel\sqrt{\overline{C}_1a_1+\overline{C}_2a_2+\sqrt{\mathcal{A}}}+\mathcal{O}[q_\parallel^2]\ ,
\nonumber\\
&& \frac{\Omega_3(q_\parallel)}{\omega_p}=q_\parallel\sqrt{\overline{C}_1a_1+\overline{C}_2a_2-\sqrt{\mathcal{A}}}+\mathcal{O}[q_\parallel^2]\ ,
\end{eqnarray}
where $\mathcal{A}\equiv(\overline{C}_1a_1-\overline{C}_2a_2)^2+4\overline{C}_1\overline{C}_2 a_1^2$ and the term $4\overline{C}_1\overline{C}_2 a_1^2$ plays the role of ``Rabi coupling''.
Clearly, for long wavelengths, only $\Omega_1(q_\parallel)$ depends on $\omega_p$.
However, the excitation spectrum changes dramatically when a current is driven
through the configuration. Under a constant electric field, the carrier distribution
is modified, as may be obtained by employing the relaxation-time approximation
in the equation of motion for the center-of-mass momentum. For a parabolic energy band for
carriers with effective mass $m^\ast$  and  drift velocity ${\bf v}_d$  determined by
the electron mobility and the external electric field, the electrons in the medium
are redistributed. This is determined by  a momentum shift in the
wave vector ${\bf k}_\parallel\to {\bf k}_\parallel-m^\ast {\bf v}_d/\hbar$ in the thermal-equilibrium energy
distribution function $f_0(\varepsilon_{\bf k})$. By making a change of variables
in the well-known Lindhard polarization function $\Pi^{(0)}(q,\omega)$, this effect is
equivalent to a frequency shift $\omega\to \omega-{\bf q}\cdot {\bf v}_d$.
For massless Dirac fermions in graphene with linear energy dispersion, this Doppler
shift in frequency is not in general valid for arbitrary wave vector. This is our
conclusion after we relate
the surface current density to the center-of-mass wave vector in a steady state.
Our calculation shows that the redistribution of electrons  leads to a shift in the
wave vector appearing in the Fermi  function by the center-of-mass wave vector
${\bf K}_0=(k_F/v_F){\bf v}_d$, where $k_F$ and $v_F$ are the Fermi wave vector and
velocity, respectively. However, in the long-wavelength limit, $q_\parallel\to 0$, the Doppler
shift in frequency is approximately obeyed. Consequently, regardless of the nature
of the 2D layer represented in the dispersion equation we may replace
$\omega\to \omega-{\bf q}\cdot {\bf v}_d$ in the dispersion equation in the presence
of an applied  electric field at long wavelengths.
\medskip

\begin{figure}[t]
\centering
\includegraphics[width=0.45\textwidth]{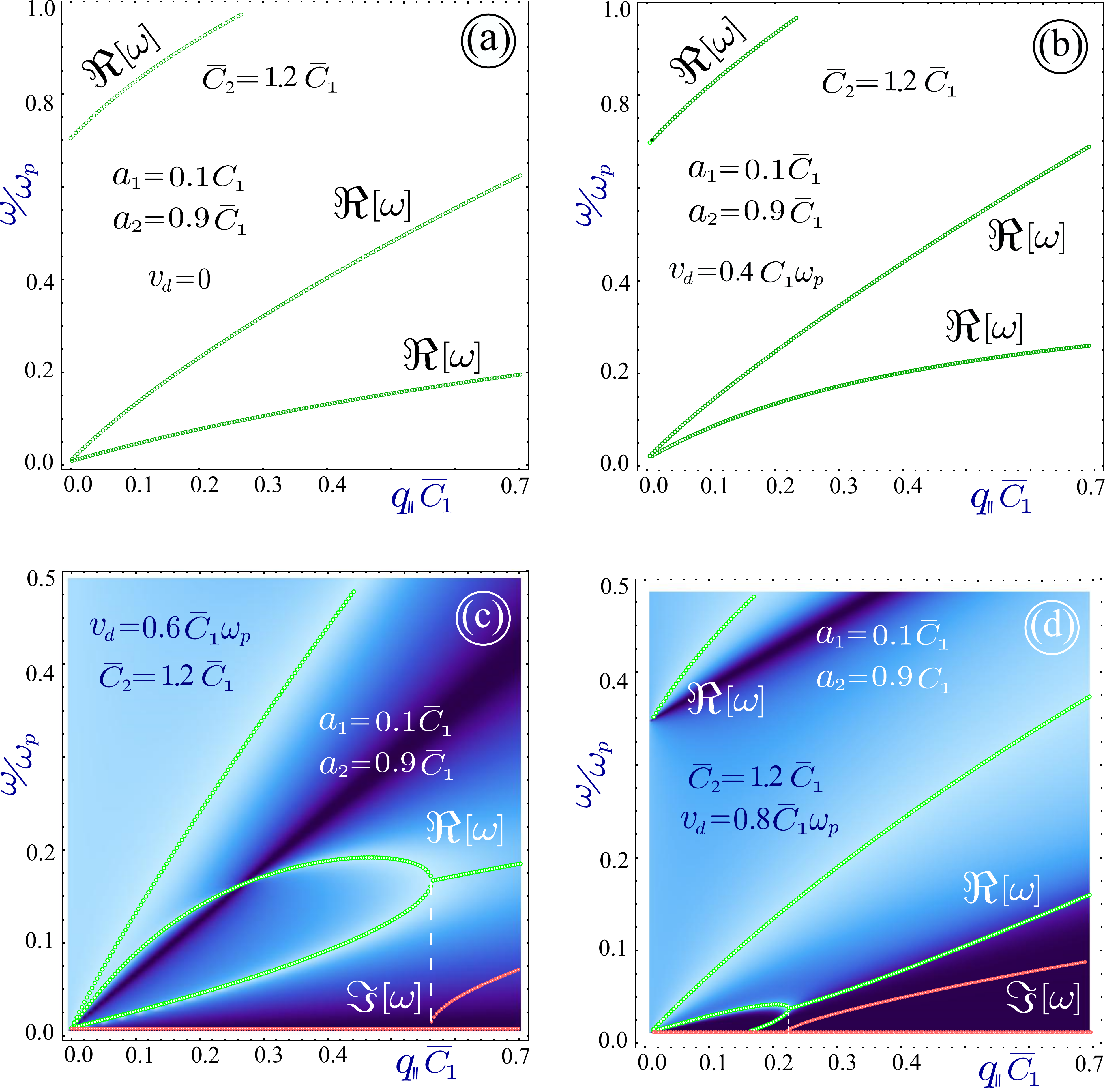}
\caption{(Color online) Plasmon dispersion relation and inverse growth
rate for a pair of 2D layers
and a semi-infinite conducting medium. The layers are located at $a_1=0.1\bar{C}_1$ and $a_2=0.9\bar{C}_1$ with respect to the surface.
Each panel is attributed to a different drift velocity with $v_d/\omega_p=0,\,0.4\bar{C}_1,\,0.6\bar{C}_1$ and $0.8\bar{C}_1$.
In panels $(a)$, $(b)$ and $(c)$, the current
is passed through the bottom 2D layer but in panel $(d)$, the current is passed through
the semi-infinite medium. The particle concentrations, temperatures or chemical
potentials for the 2D layers is such that $\bar{C}_2=1.2\bar{C}_1$ for all
cases. Only the lowest plasmon branch becomes unstable beyond a critical
wave vector $q_c$. The Rabi-type splitting of the plasmon
excitation branches by the external electric field indicates the excitation
of two quasiparticles with different excitation energies at the same wavelength.}
\label{FIG:3}
\end{figure}
To highlight the effect due to a surface in spiler, as a comparison we first present in Figs.\,\ref{FIG:1}(a)
and \ref{FIG:1}(b) the plasmon dispersion for an isolated pair of 2D layers in the absence\,\cite{DasSarma}
and presence\,\cite{Bakshi} of a current, respectively. When a substrate
surface plasmon is not contributing, the plasmon instability starts at $q_\parallel=0$
and exists over a finite range until a bifurcation point is reached for $q_\parallel$.
By going beyond this bifurcation point, the interlayer Coulomb coupling is effectively suppressed, leading to one uncoupled 2D-sheet plasma
and two current-split 2D-sheet plasmon modes.
However, when a surface plasmon interacts with the two 2D layers,
the plasmon instability may be moved to shorter wavelengths, as we clearly illustrated in Figs.\,\ref{FIG:1}(d), where
we show the results when spiler consists of a 2D layer and a semi-infinite conducting medium.
Plasmon remains stable in a single current-driven 2D layer.
In the presence of the surface plasmon, as $v_d/\omega_p=0.6\bar{C}_1$ in $(d)$ the plasmon instability occurs outside the closed Rabi ``loop'', i.e., $q_\parallel>q_c$ ($q_c$ is the critical wave vector),
and at shorter wavelengths by pushing $q_c$ significantly above zero.
\medskip

\begin{figure}
\centering
\includegraphics[width=0.35\textwidth]{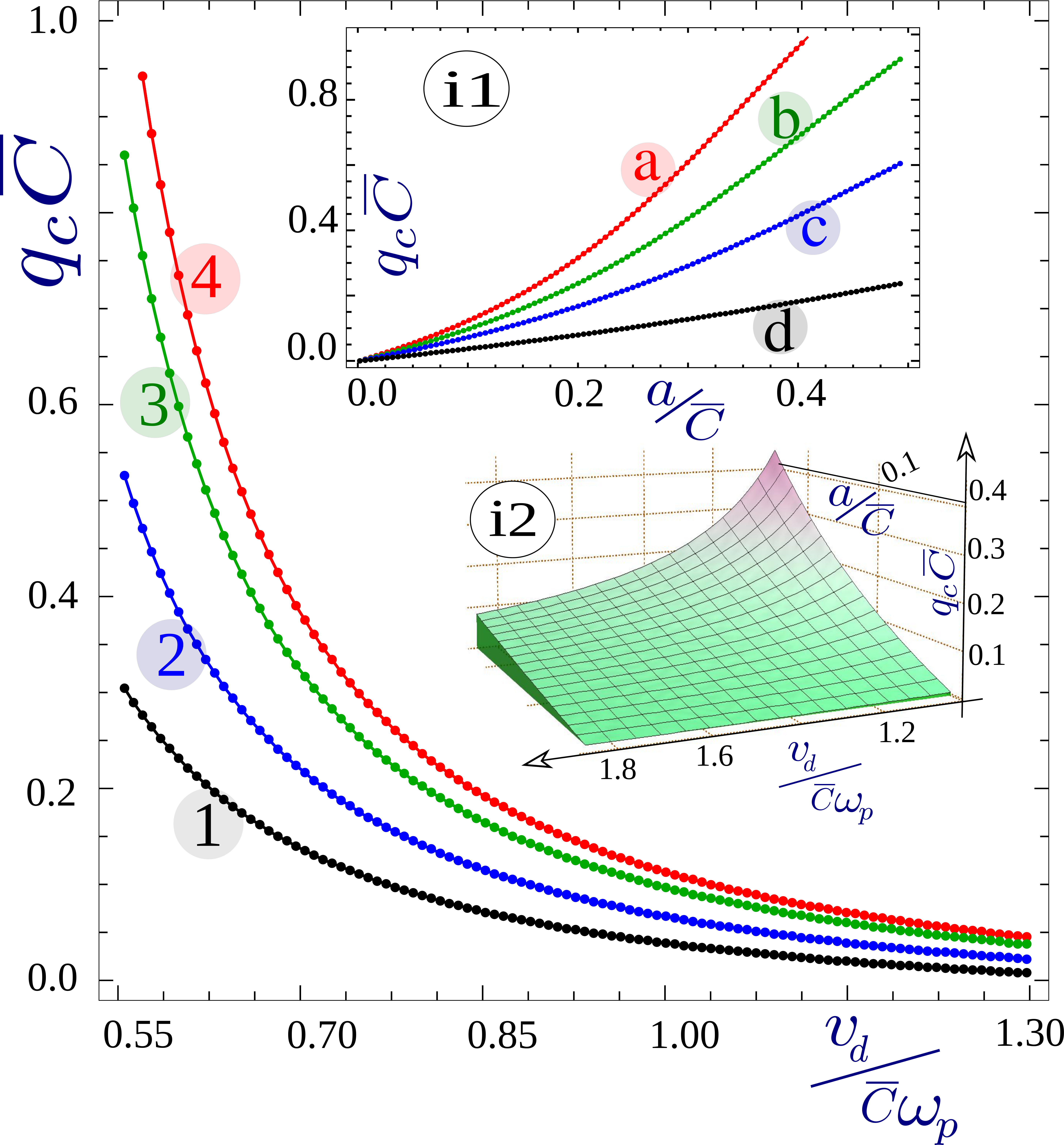}
\caption{(Color online) Plot of the the critical wave vector where the plasmon instability occurs
for a single 2D layer Coulomb coupled to a thick conducting medium as a function
of the drift velocity $v_d$.
The curves $1$-$4$ correspond to chosen separations from the surface with $a/\overline{C}=1.1$, $1.0$,
$0.8$ and $0.6$. Inset $i1$ shows the variation of $q_c$ with $a$. Curves
$a$-$d$ correspond to $v_d=1.0$, $1.10$, $1.25$ and $1.7$ in units of $\bar{C}\omega_p$.
Inset $i2$ is a 3D plot of $q_c$ as functions of both $a$ and $v_d$.}
\label{FIG:4}
\end{figure}
In order to get a full understanding of the mechanism for instability shown in Fig.\,\ref{FIG:1}, in Fig.\,\ref{FIG:2} we have numerically investigated  the effect of a passing current through a layer of
2DEG, graphene or silicene which is Coulomb coupled to a conductor. Specifically, we consider a pair of 2D layers
and a semi-infinite  medium such as a heavily-doped semiconductor. We present in Fig.\,\ref{FIG:2} both the plasmon dispersion and the inverse growth rate.
Each panel shows results for a different drift velocity given by $v_d/\omega_p=0,\,0.5\bar{C}_1,\,0.8\bar{C}_1$
and $1.6\bar{C}_1$. In the absence of a current, panel $(a)$ shows
that there are three plasmon excitation branches, which are stable as given by
Eq.\,(\ref{solutions}). At low $v_d$, panel $(b)$ demonstrates that the
plasmons are still stable. However, as $v_d$ is increased further, the lowest branches
may become unstable as in $(c)$ and $(d)$ through the appearance of a positive
imaginary part for the frequency.
There is a threshold value for $v_d$ beyond which the plasmon excitation becomes unstable. On the other hand,
the existence of the surface plasmon greatly screens both the interlayer and intralayer Coulomb couplings as $q_\parallel a_1\ll 1$ in the range of $\omega/\omega_p\ll 1$.
This stabilizes the plasmon excitation for $q_\parallel<q_c$ by suppressing the interlayer coupling as shown in $(c)$.
As $v_d$ increases to $1.6\bar{C}_1\omega_p$ in $(d)$, $q_c$ reduces almost to zero, but the wide stable region in $(c)$ is squeezed into a narrow belt.
The occurrence of such a new unstable region starting from $q_\parallel=0$ is a combined result from both the surface-induced softening of the two 2D-sheet plasma modes to two acoustic-like plasmon modes
as well as the strong interlayer coupling for small layer separation. To some extent, this feature is similar to the result displayed in Fig.\,\ref{FIG:1}$(b)$ for the isolated pair of 2D layers.
\medskip

Figure\ \ref{FIG:3} illustrates our results for larger 2D layer separations
from each other compared to the case in Fig.\,\ref{FIG:2}. Panel $(a)$ again corresponds to
the analytic solutions in Eq.\,(\ref{solutions}) when $v_d=0$. The plasmon excitations are all still
stable in $(b)$ at $v_d/\omega_p=0.4\overline{C}_1$. However, as
$v_d$ is increased further in $(c)$ and $(d)$, an instability appears at $q_c$, which is exactly where the Rabi ``loop'' for plasmon excitations closes in the two lower panels.
Beyond $q_c$, the plasmon becomes unstable corresponding to opposite phase velocities for two current-split plasmon branches.
The loop shape is quite different for current passing through the 2D layer [in $(c)$] and the semi-infinite medium [in $(d)$]. We also note
that by adjusting the layer separation, we may change $q_c$ value for controlling the onset of the plasmon instability.
With such a large interlayer separation, the interlayer interaction becomes very weak and the system effectively
behaves like the single current-driven 2D layer coupled to a conducting surface, similar to that
in Figs.\,\ref{FIG:1}$(d)$.
\medskip

Physically, from the point of view of momentum space, electrons may only occupy   momentum
space within the range of $|{\bf k}_\parallel|\leq k_F$ at zero temperature in a state of thermal
equilibrium, where $k_F$ is the electron Fermi wave number and
$\varepsilon(k_F)=\varepsilon(-k_F)=E_F$ is the Fermi
energy. When a current is passed through the electron gas, electrons are driven out
from this thermal-equilibrium state and their population becomes asymmetrical
with respect to $k_\parallel=0$. In this case, the Fermi energy $E_F$ is split into
$E_{F,+}=\varepsilon(k_F+K_0)$ and $E_{F,-}=\varepsilon(-k_F+K_0)$ with $E_{F,+}>E_{F,-}$,
where $\hbar K_0$ represents the electron center-of-mass momentum. In this shifted
Fermi-Dirac distribution model, electrons in such a non-equilibrium state are
energetically unstable, and the higher-energy electrons in the range
$k_F\leq k\leq k_F+K_0$ tend to decay into lower-energy empty states by emitting
EM waves and phonons to ensure the conservations of total momentum and energy.
\medskip

The current-driven asymmetric electron distribution in ${\bf k}_\parallel$ space leads to an induced oscillating
polarization current or a ``dipole radiator''. If two electron gas layers are
placed  close enough, the in-phase interlayer Coulomb interaction will give rise to a
dipole-like plasmon excitation, similar to that of a single layer. On the other hand,
the out-of-phase interlayer Coulomb coupling will lead to a quadruple-like excitation.
This quadruple-like plasmon excitation can be effectively converted into a transverse
EM field in   free space if a surface grating is employed.
\medskip

The surface-induced instability in spiler may lead to EM radiation,
and the lower edge of its radiation frequency can be tuned directly by $q_c$ to cover the THz frequency range.
Here, we present in Fig.\,\ref{FIG:4} the dependence of controlling parameter $q_c$ as functions of $v_d$ and separation
$a$ for a single 2D layer coupled to surface plasmon, where $q_c$ is increased with either reducing $v_d$ or increasing $a$.
These results clearly demonstrate significant shifts of $q_c$
within the desired ranges for operations of both photodetectors and EM-wave devices.
\medskip

In summary, we are proposing a spiler quantum plasmonic device which employs
2D layers such as graphene/III-V semiconductor
hybrids in combination with a thick conducting material with clean interfaces. We find that the spiler spontaneously emits
EM radiation when a current is passed through the 2D layer or the underlying conductor to make the plasmons become unstable at a
specific frequency and wave number. It is possible to tune the onset of plasmon
instability by selecting the properties of the nanosheet or frequency of the
surface plasmon, i.e., the substrate. The surface plasmon plays a crucial role
in giving rise to the splitting and the concomitant streams of quasiparticles
whose phase velocities are in opposite directions when the instability takes place.
The emitted EM radiation may be coupled easily outward to a free space by a grating on the surface.
Finally, we note that in  presenting our numerical results, we  measured
frequency  in terms of the bulk plasmon frequency which, typically for conductors, is
$\hbar\omega_p\sim  0.5$\,eV. Either for intrinsic graphene, doped monolayer graphene
or an inversion layer 2DEG,  we have $\overline{C}\sim 10^{-7}$\,m and $v_d\sim 10^6$\,m/s.
The frequency unit used in Fig.\,\ref{FIG:1} is $\omega_0=\sqrt{2\pi e^2 C_1 k_F/\epsilon_s}$
which is of the same order as $\omega_p$.
\medskip

Experimentally, the emitted EM radiation of a spiler device can be detected by heterodyne mixing technique using a planar Schottky diode\,\cite{reno}.
In this experiment, the mixing of the EM radiation against a molecular laser line is known to give a high precision measurement of the EM radiation
frequency and can also show transient turn-on behavior in a pulsed spiler device.
\medskip

The spontaneous EM radiation from the proposed spiler does not require a population inversion in a laser, a coupling-field quantum coherence in amplification without inversion and
a exciton-polariton condensation\,\cite{laser,LWI,polar}.
Instead, it depends on energy conversion\,\cite{spaser} from an applied dc electric field to an optical field based on a current-induced plasmon instability.
The radiation frequency of the splier is tunable and covers the whole THz range.
Terahertz waves are able to penetrate materials that block visible light and have a wide range of possible applications, including chemical analysis, security scanning, medical imaging, and telecommunications.
\medskip

\acknowledgments
This research was supported by  contract \# FA 9453-13-1-0291 of AFRL. DH would like
to thank the Air Force Office of Scientific Research (AFOSR) for its support.
We thank Oleksiy Roslyak and Antonios Balassis for helpful discussions. 
WP was supported by the U.S. Department of Energy, Office of Science, Basic Energy Sciences, Materials Sciences and Engineering Division.

\end{document}